**Energy-Dependent Dechanneling in Cu: Insights from Monte Carlo Channeling Simulations**

*Przemysław Jóźwik*, Cyprian Mieszczyński, Renata Ratajczak,* Andrzej Turos

P. Jóźwik, C. Mieszczyński, R. Ratajczak, A. Turos
National Centre for Nuclear Research, A. Soltana 7, 05-400 Otwock, Poland
E-mail: przemyslaw.jozwik@ncbj.gov.pl

Ion channeling and backscattering techniques are powerful tools for studying crystal lattice disorders and defect structures in crystalline materials. However, the accurate interpretation of channeling phenomena necessitates the utilization of simulation models that account for the intricate interactions between point defects, dislocations, and extended defect clusters. The present paper introduces a Monte Carlo method that reproduces experimental spectra over a wide range of analyzing beam energies and enables quantitative identification of defect types and distributions. The simulations reveal characteristic energy dependencies that distinguish point defects from extended defects, offering a novel perspective on disturbances caused, for example, by ion implantation in metals and semiconductors. To this end, the McChasy code has been developed as a flexible and accessible tool for scientists, enabling the modeling of various crystal systems, including complex semiconductors, multilayer epitaxial films, and oxide crystals. The program's integration of experimental data on ion channeling with defect modeling establishes a robust framework for defect analysis in materials science. The present article expounds upon the simulation capabilities of the program by reproducing the characteristic "elbows" in channeling spectra that were previously observed in experiments conducted on Cu crystals.

## 1. Introduction

The damage caused by radiation in crystalline metals is primarily characterized by the formation of vacancies and interstitial atoms. These atoms, under the influence of increasing radiation fluences and as a result of migration and agglomeration, can transform into more complex structures, such as amorphous clusters or dislocation loops. The identification of the nature of these defect clusters, whether voids or interstitial clusters, and whether dislocation-type or amorphous, remains a significant challenge in materials science. This is due to the direct impact on the mechanical stability and electronic properties of materials exposed to radiation. Conventionally, transmission electron microscopy (TEM) has been employed to characterize such defects; however, complementary spectroscopic methods provide valuable insights into the underlying mechanisms. A notable advantage of these methods is their ability to facilitate experiments and collect responses from the entire depth of the sample.
Rutherford backscattering spectrometry (RBS) in conjunction with ion channeling (RBS/C) has been demonstrated to be a reliable technique for the study of disorder in irradiated crystals[1]. Specifically, the analysis of dechanneling behavior as a function of probe ion energy enables the differentiation of various defect types. Agrawal and Sood[2] demonstrated that the channeling spectra of self-implanted copper exhibit characteristic "knees" in the dechanneled fraction derived from backscattering yield, and the cross section of dechanneling demonstrates an energy dependence consistent with interstitial atoms. These observations were partly consistent with TEM studies that revealed interstitial dislocation loops in heavily irradiated copper.
In this study, we are building upon this line of research by employing Monte Carlo (MC) simulations[3] of channeling spectra in copper (Cu). Our simulations successfully replicate the manifestation of “knees” in smoothed spectra at comparable depths and energies, thereby substantiating the hypothesis that inter-nodal defect clusters function as predominant dechanneling centers. The objective of this study is to establish a quantitative framework that links experimental RBS/channeling results with computational modeling through systematic analysis of the energy dependence of simulated dechaneling cross sections. This approach has the potential to not only corroborate the findings of experimental studies but also to augment the capacity to predict the characteristics of defects in irradiated metals.
In order to investigate the influence of defect morphology, a uniform distribution of defects within 400 nm of the surface was utilized in several configurations. In addition to differentiating between edge dislocations and dislocation loops, the study also examined the variation in size of dislocation loops. This facilitated a systematic investigation into the dependence of channeling spectra on the dimensions of defect clusters and their spatial distribution. Moreover, the necessity of determining the geometric parameters of the tested structure was demonstrated by showing differences in the intensity of simulation spectra obtained for different parameters.

**2. Methods**

Monte Carlo simulations of the ion channeling process were performed using the McChasy code[3–5], which is widely used for modeling direct scattering and defect-induced dechanneling in ion-bombarded/implanted crystalline materials. The focus of the simulations was on copper (Cu) single crystals oriented along the ⟨110⟩ axis. This was done to reproduce the geometry that was utilized in the experimental studies published in Agrawal et al.[2] In these studies, self-implanted copper single crystals were investigated. The RBS experiment parameters provided by Agrawal et al. were kept for MC simulations, i.e., backscattering angle of 164°, He beam energies of 2.0, 2.9, and 3.5 MeV, <001> orientation of Cu crystal and integrated helium dose of 15 µm. The parameters of the experiment that were not specified in Ref. [2] were adopted ad hoc based on the values used in similar studies. The following are included: energy resolution: 20 keV, ion-beam dispersion: 0.03°, and thickness of samples: 1150 nm.

A series of simulations were conducted to assess the effectiveness of the proposed approach under various conditions as described in Table 1. Defect structures were introduced into the simulation cell based on established models of edge dislocations[6,7] (DIS) and dislocation loops (DLPs),[3] as described in detail elsewhere. These models are characterized by their consideration of the deformation field, which refers to the decay of atomic plane bending (that follows the arctan function according to the Paierls-Nabarro model[8–10]) with distance from the defect. This feature enables a realistic reproduction of the dechanneling phenomenon. For all simulations, the DIS and DLP profiles were considered constant, with densities of $2x10^{10}$ $cm^{-2}$ and spreading up to a depth of 400 nm.

*Table 1 Simulation conditions - applied models of defects and selected parameters*

| Applied marking | Simulation condition |
|---|---|
| **DLP 10-25** | Dislocation loops with random diameters ranging from 10 to 25 nm |
| **DLP 15** | Dislocation loops with a fixed diameter of 15 nm |
| **DLP 5-10** | Dislocation loops with random diameters ranging from 5 to 10 nm |
| **DIS** | Edge dislocations |
| **DLP 5-10 (ZnO)** | Dislocation loops with random diameters ranging from 5 to 10 nm but with geometric parameters obtained for ZnO |

With the exception of the simulations designated as DLP 5-10 (ZnO), all others were executed with consideration for the geometric parameters of dislocations ascertained for SrTiO3,[7] given the unavailability of parameters for Cu and the cubic nature of both structures. The objective of this work is to present the computational capabilities of the McChasy program, rather than to directly compare the simulation results with experiment. Therefore, the aforementioned simplification is deemed appropriate. However, it should be noted that prior to conducting the analysis of experimental RBS spectra obtained for Cu using the McChasy program, the geometric parameters of dislocations for this structure must be determined. This determination can be made using HRTEM[5] or MD[11].

The simulated spectra were analyzed in the mode of the ratio of aligned to random spectra, with particular attention paid to the appearance of characteristic "knees," which are identified with the boundary of the defective area. By comparing the dependence of the cross-section energy of dechanneling for different defect configurations, the simulations provide a quantitative framework for interpreting experimental observations in irradiated copper and related metals.

**3. Results and discussion**

The simulated spectra for all conditions are displayed in Figure 1. The spectrum called “random“ was calculated for a tilt of a sample at an angle of 4° and rotating it around the axis. All other spectra are so called aligned spectra and refer to a scenario wherein a He beam is bombarded onto a crystal along the orientation direction, specifically the <001> orientation in this particular instance. The term "pristine" is used to describe a "perfect" crystal that does not contain intentionally introduced defects. In contrast, other aligned spectra refer to samples containing defects, typically resulting from ion implantation.

As a consequence of interaction with target atoms (Cu), He ions undergo a loss of energy as they penetrate deeper into the structure. Ions with a sufficiently low impact parameter undergo backscattering on Cu nuclei and lose a certain amount of energy resulting from the kinematic factor, which for a given geometry (i.e., for a scattering angle of 164°) is 0.7808. The energies of He ions upon reaching the sample surface and corresponding energies after backscattering on surface atoms are shown in Table 2. The formation of well-pronounced surface peaks is evident at these energies, as evidenced by the aligned spectra in Figure 1. Table 2 also includes the energies of the beam upon reaching a depth of 400 nm, then after colliding with Cu atoms at that depth, and eventually upon passing through the surface on its way back to the detector (i.e., the detection energy that is visible in the RBS spectrum). The energy values resulting from energy losses were calculated using the He ion energy loss tables in Cu obtained from SRIM.[12]

In the context of backscattering on a surface, the detection energy is equivalent to the energy that remains following the collision. This is due to the absence of energy loss resulting from interaction with target atoms. The dashed vertical lines illustrated in Figure 1 (and Figure 2) correspond to detection energies that are indicative of backscattering events occurring on the surface and at a depth of 400 nm.

*Table 2. Energy values before and after collision with Cu atoms leading to backscattering on the surface and at a depth of 400 nm, and detection energies.*

| Depth | He-ion energy [keV] | | | |
|---|---|---|---|---|
| | (incident) | (before collision) | (after collision) | (detected) |
| **Surface** | 3500 | 3500 | 2733 | 2733 |
| **400 nm** | | 3313 | 2587 | 2374 |
| **Surface** | 2900 | 2900 | 2264 | 2264 |
| **400 nm** | | 2698 | 2106 | 1879 |
| **Surface** | 2000 | 2000 | 1562 | 1562 |
| **400 nm** | | 1769 | 1382 | 1134 |

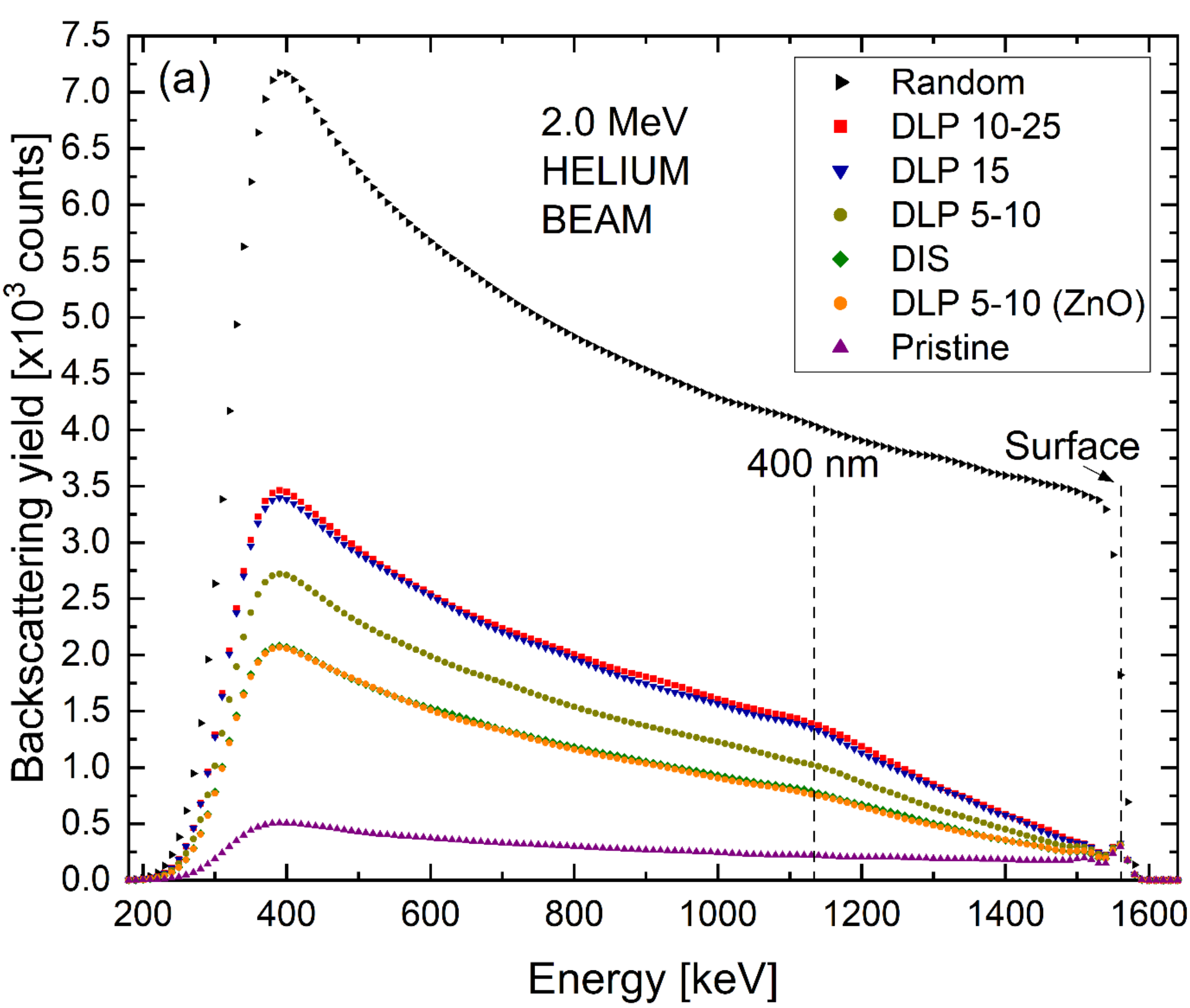
(a)
2.0 MeV
HELIUM
BEAM
Random
DLP 10-25
DLP 15
DLP 5-10
DIS
DLP 5-10 (ZnO)
Pristine
400 nm
Surface
Backscattering yield [x10$^3$ counts]
Energy [keV]
7.5
7.0
6.5
6.0
5.5
5.0
4.5
4.0
3.5
3.0
2.5
2.0
1.5
1.0
0.5
0.0
200
400
600
800
1000
1200
1400
1600

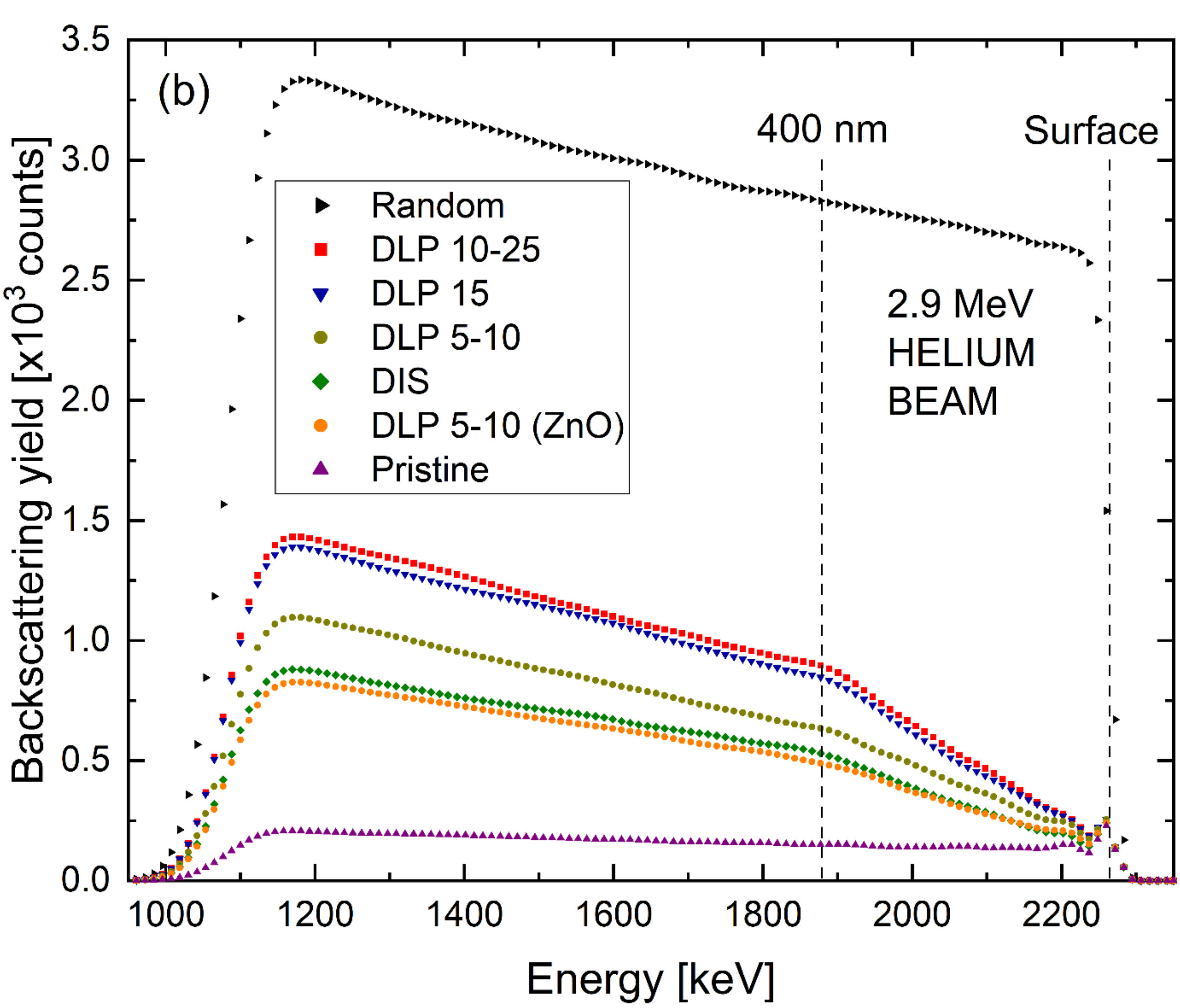
(b)
400 nm
Surface
Random
DLP 10-25
DLP 15
DLP 5-10
DIS
DLP 5-10 (ZnO)
Pristine
2.9 MeV
HELIUM
BEAM
Backscattering yield [x10$^3$ counts]
Energy [keV]

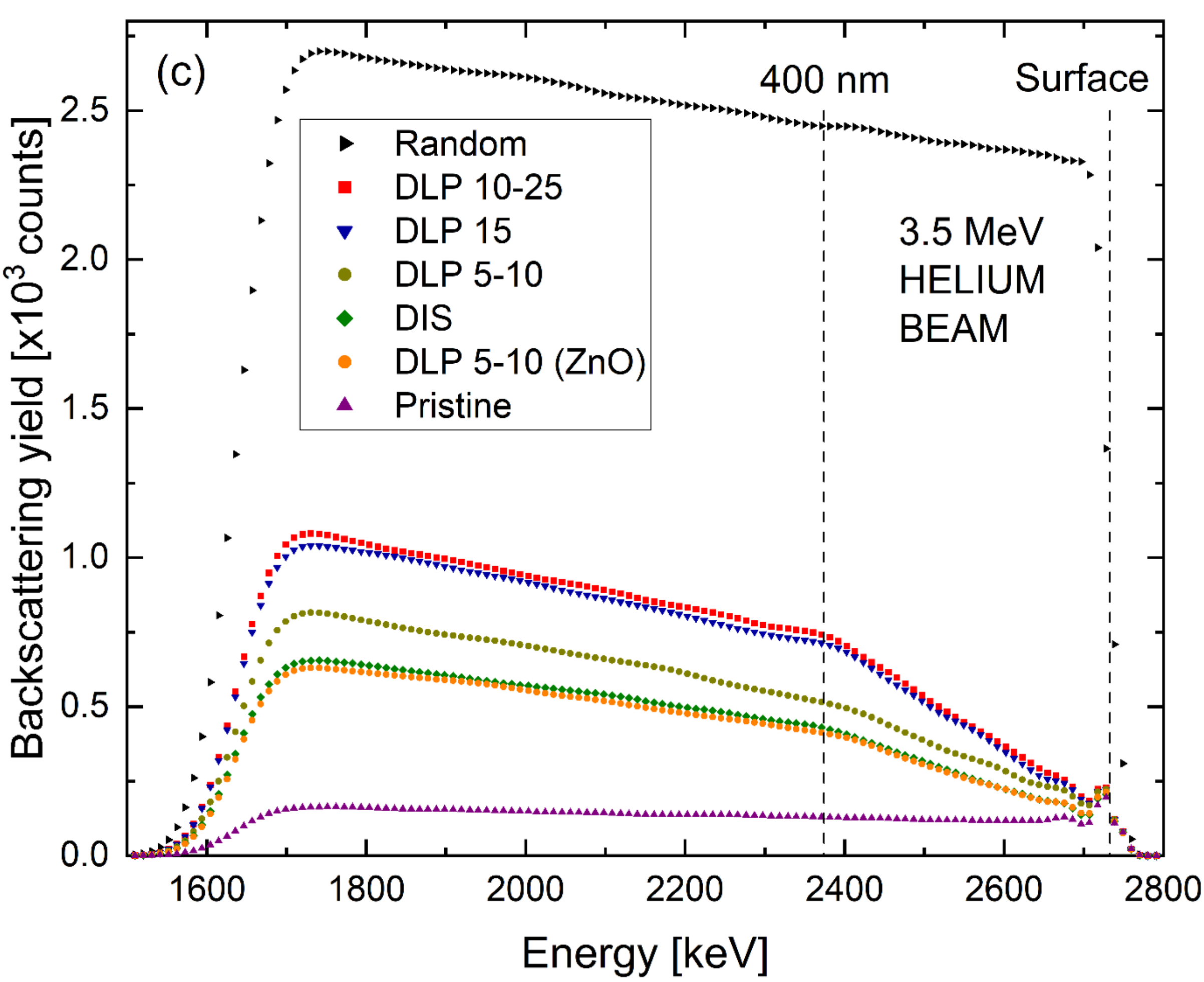

Error! Reference source not found. RBS/C simulated spectra for <011> Cu as obtained from the McChasy code for incident He-ions of different energies: (a) 2.0 MeV, (b) 2.9 MeV, and (c) 3.5 MeV. Spectra were calculated for different defect types and parameters, as described in Table 1, of constant density of $2x10^{10}$ $cm^{-2}$ spreading up to a depth of 400 nm.

Adapting the convention utilized by Agrawal et al., Figure 2 illustrates the relative dechanneled fraction, which is calculated as the ratio of the aligned spectrum to the random spectrum. The formation of "knees" exhibits a strong correlation with a depth of 400 nm, which delineates the boundary between defective and non-defective regions. The observation of this phenomenon was initially made through experimental means by Agrawal et al. Subsequent validation has now been achieved through simulations. This finding lends further credence to the efficacy of the extended defect models developed for the McChasy program, in conjunction with the previously published analysis of RBS measurements at multiple beam energies in ion-implanted GaN.[13]

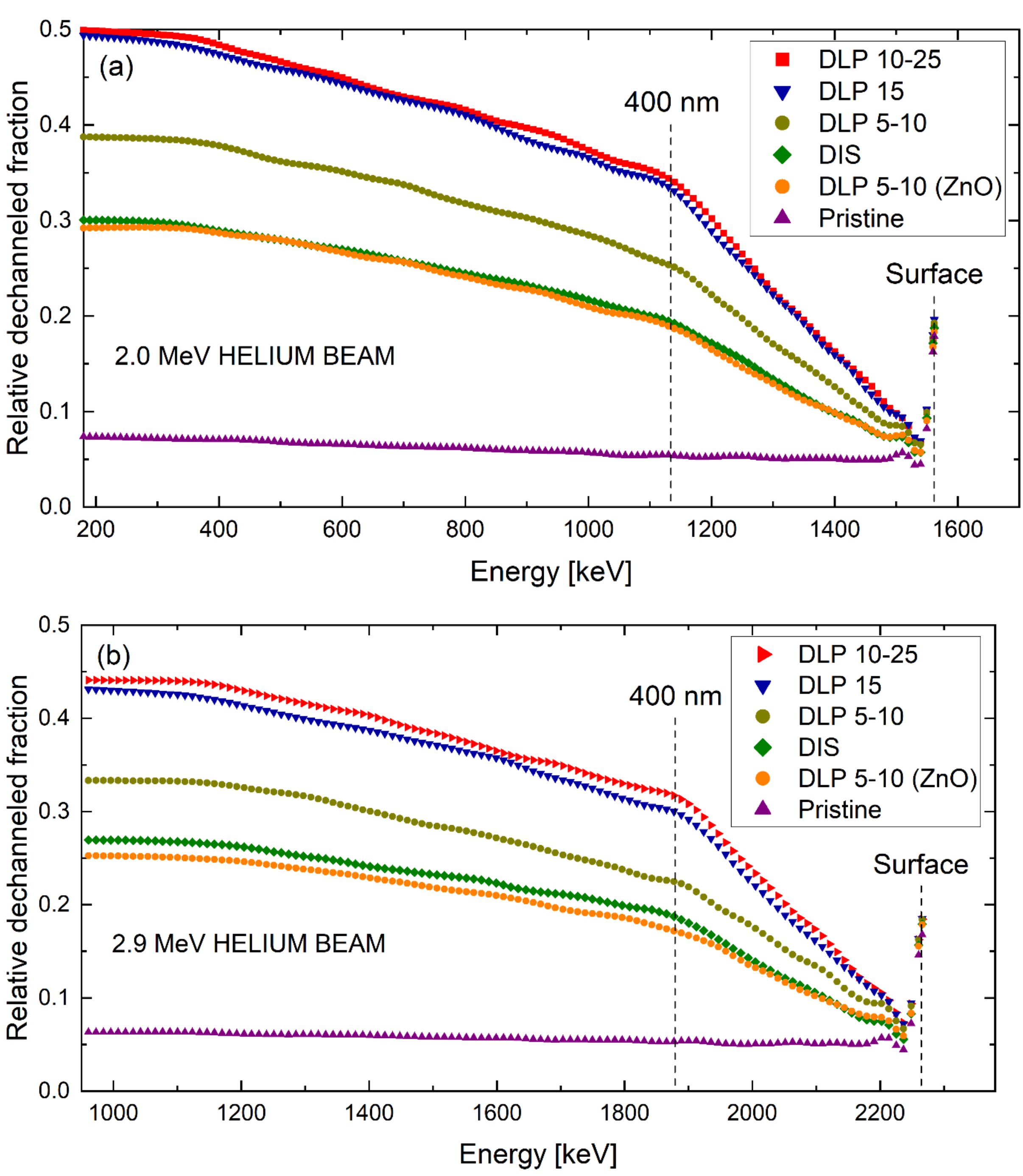
(a)
DLP 10-25
DLP 15
DLP 5-10
DIS
DLP 5-10 (ZnO)
Pristine
400 nm
Surface
2.0 MeV HELIUM BEAM
Relative dechanneled fraction
Energy [keV]
(b)
400 nm
Surface
2.9 MeV HELIUM BEAM

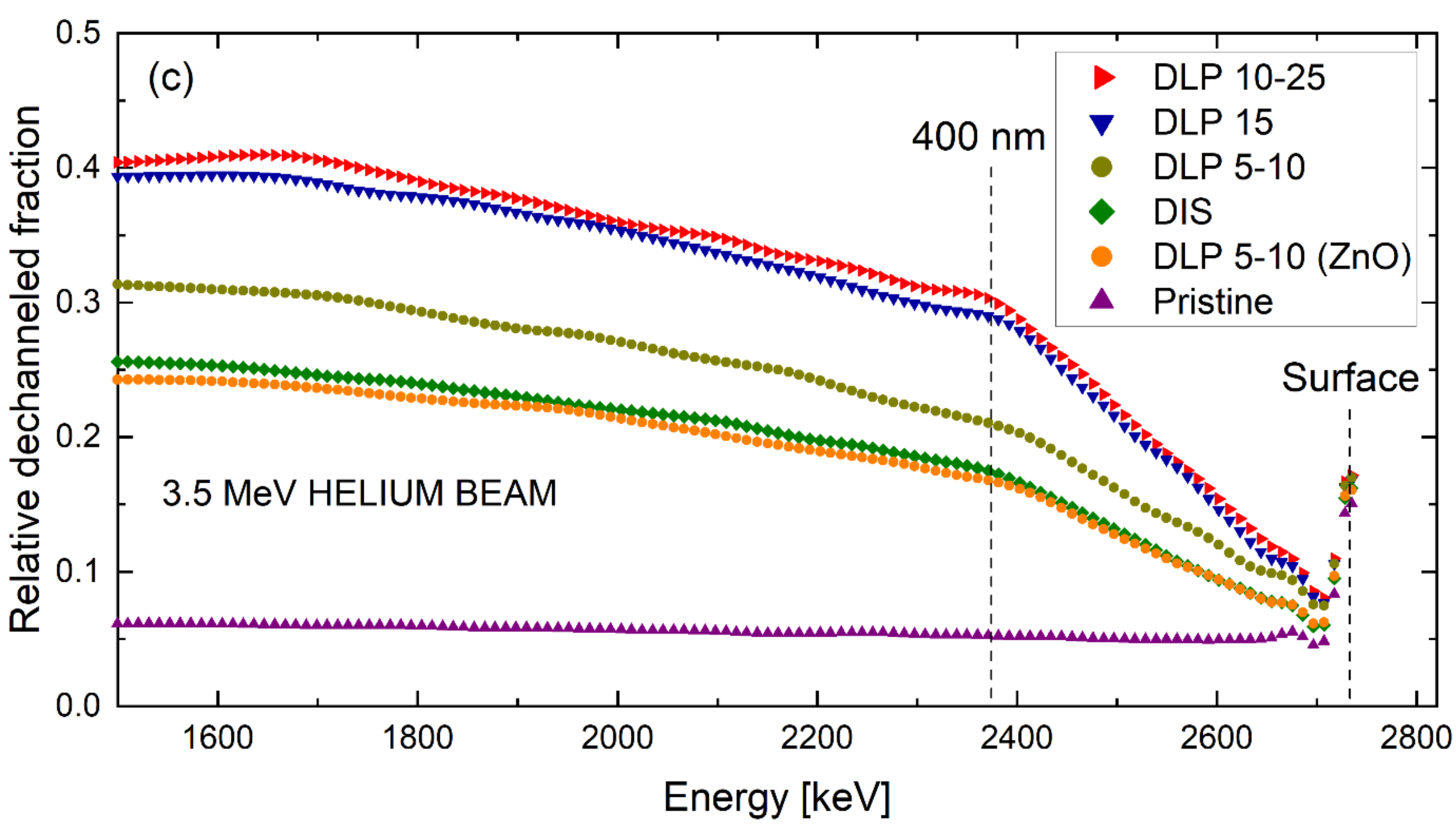

**Figure 2.** Relative dechanneled fraction as extracted from RBS/C spectra shown in Figure 1 for incident He-ions of different energies: (a) 2.0 MeV, (b) 2.9 MeV, and (c) 3.5 MeV. Spectra were calculated for different defect types and parameters, as described in Table 1, of constant density of $2x10^{10}$ $cm^{-2}$ spreading up to a depth of 400 nm..

A comparison of the results obtained for different simulation variants revealed clear differences in spectrum intensity. All spectra are characterized by a smoothly increasing dechanneling signal with varying slopes, depending on the defect model and the parameters utilized. In the case of only edge dislocations, a relatively low signal is observed. Conversely, simulations incorporating dislocation loops tend to yield a higher spectrum intensity. This phenomenon can be attributed to the increased distortion of the structural elements surrounding the loops as opposed to the dislocations themselves. Additionally, larger-diameter loops have been shown to generate a higher level of dechanneling, which is associated with an increase in structural distortion as the size of the loop increases. It was observed that only simulations with random loop sizes between 5 and 10 nm yielded a lower spectrum in comparison to simulations incorporating edge dislocations.

As previously stated, the spectra were obtained for geometric parameters derived for $SrTiO_3$. The significance of these parameters in simulations has previously been demonstrated for Al.[5] Furthermore, simulations have been conducted employing parameters that have been determined for ZnO. The intensity of the spectrum has evidently diminished, as evidenced by the observation of disparate deformation fields in the vicinity of dislocations/dislocation loops in both structures. These fields are distinguished by variations in the degree of bending of atomic planes and the subsequent dissipation of this bending. Consequently, these structures contribute disparately to the dechanneling of the beam ions. The employment of disparate simulation variants enables not only a superior alignment of the simulation with experimental data but also the differentiation of the predominant defect types in the material under study.

**4. Conclusion**

The development and application of the McChasy Monte Carlo channeling code has demonstrated its effectiveness in reproducing ion backscattering spectra over a wide range of conditions that mimic experimental conditions. The incorporation of sophisticated models of extended defects, including dislocations and dislocation loops, has empowered the program to offer a malleable framework for interpreting intricate dechanneling phenomena in crystalline materials.

The following key simulation results in copper have been identified:

- The reproducibility of experimental channeling spectra for both ideal and defect-containing crystals has been demonstrated.
- The ability to distinguish between different types of defects based on their characteristic energy dependence and depth profiles is paramount. In addition to the extensive array of defects that have been examined in this study, the program has been developed to consider randomly displaced atoms as well.
- The enhanced precision in defect quantization during simulations conducted at multiple analytical beam energies has been demonstrated to mitigate ambiguities in experimental interpretation.

The practical usability of Monte Carlo channeling analysis on standard personal computers has enabled researchers to incorporate this analysis into routine studies of semiconductor compounds, multilayer epitaxial films, and oxide crystals.

In summary, McChasy broadens the scope of ion channel analysis beyond conventional analytical approaches by providing a robust simulation environment that integrates experimental data and defect modeling. Subsequent enhancements will center on augmenting the defect structure library, refining crystal lattice distortion models, and integrating with complementary characterization techniques to furnish a comprehensive toolkit for defect analysis in advanced materials. The program is available at no cost to the user, subsequent to completion of the requisite form on the program's website.[14]

**Author Contributions**

P. Jóźwik: Conceptualization (equal); Data Curation (lead); Formal analysis (lead); Investigation (lead); Methodology (equal); Resources (equal); Software (lead); Supervision (equal); Validation (lead); Writing - original draft (lead); Writing - review & editing (equal); Visualization (lead); C. Mieszczyński: Conceptualization (equal); Investigation (equal); Supervision (equal); Validation (equal); Writing - review & editing (equal); R. Ratajczak: Investigation (equal); Supervision (equal); Validation (equal); Writing - review & editing (equal); A. Turos: Investigation (equal); Supervision (equal); Validation (equal); Writing - review & editing (equal).

**Acknowledgements**

This research was supported by statutory research funds. The authors also acknowledge the use of DeepL Write software for assistance in language corrections and stylistic improvements.

Received: ((will be filled in by the editorial staff))
Revised: ((will be filled in by the editorial staff))
Published online: ((will be filled in by the editorial staff))